\documentclass[a4paper]{jpconf}
\usepackage{graphicx}

\def\pr{\mbox{PG~2131+066}}  
\def\pt{\mbox{PG~1707+427}}  
  
\def\v4334{\mbox{V4334 Sgr}}

\begin{document}  
\title{Seismological  constraints   on  the  high-gravity   DOV  stars
       PG2131+066 and PG 1707+427}
  
\author{A H C\'orsico$^1$, 
        L G Althaus$^1$, 
        M M Miller Bertolami$^1$ and \\
        E Garc\'\i a--Berro$^{2,3}$}  
\address{$^1$Facultad de Ciencias Astron\'omicas y Geof\'{\i}sicas,   
         Universidad Nacional de La Plata, 
         Paseo del Bosque S/N, (1900)   
         La Plata, 
         Argentina}  
\address{$^2$Departament de F\'\i sica Aplicada, 
         Escola Polit\`ecnica Superior de Castelldefels, 
         Universitat Polit\`ecnica de Catalunya,    
         Av. del Canal Ol\'\i mpic, s/n, 
         08860 Castelldefels, 
         Spain}  
\address{$^3$Institute for Space Studies of Catalonia,
         c/Gran Capit\`a 2--4, Edif. Nexus 104,   
         08034  Barcelona,  Spain}
\ead{acorsico@fcaglp.unlp.edu.ar}  
  
\begin{abstract}  
A seismological study of the  pulsating PG1159 stars PG2131+066 and PG
1707+427 is presented.  We perform extensive adiabatic computations of
$g$-mode pulsation periods of  PG1159 evolutionary models with stellar
masses ranging  from $0.530$ to $0.741\, M_{\odot}$.  We constrain the
stellar mass of  PG2131+066 and PG 1707+427 by  comparing the observed
period  spacing of each  star with  the theoretical  asymptotic period
spacings and  with the  average of the  computed period  spacings.  We
also  employ the  individual observed  periods to  find representative
seismological models for both stars.
\end{abstract}  
  
\section{Introduction}  
  
Pulsating PG1159 stars (also called GW  Vir or DOV stars) are very hot
hydrogen-deficient  post-Asymptotic Giant  Branch  stars with  surface
layers  rich  in  He, C  and  O  (Werner  \& Herwig  2006)  exhibiting
multimode pulsations (non-radial  $g$-modes) with periods ranging from
5 to  50 minutes  (Winget \& Kepler  2008).  It is  generally accepted
that these stars have their  origin in a born-again episode induced by
a  post-AGB  He  thermal  pulse  (Althaus et  al.   2005).   Recently,
considerable observational effort has been invested to study pulsating
PG1159  stars. Particularly noteworthy  are the  observational efforts
devoted to RXJ 2117.1+3412 (Vauclair  et al. 2002), PG 0122+200 (Fu et
al. 2007) and PG 1159$-$035  (Costa et al.  2008).  On the theoretical
front, recent  important progress in the numerical  modeling of PG1159
stars  (Althaus et  al. 2005;  Miller Bertolami  \& Althaus  2006) has
paved  the  way for  unprecedented  seismological  inferences for  the
mentioned  stars (C\'orsico  et al.  2007a; C\'orsico  et  al.  2007b;
C\'orsico et al. 2008).  Here, we present a seismological study of the
high-gravity pulsating  PG1159 stars PG 2131+066 and  PG 1707+427, two
stars  that have  been intensively  scrutinized with  the  Whole Earth
Telescope.
  
Our pulsation analysis is based  on a new generation of stellar models
that  take into account  the complete  evolution of  PG1159 progenitor
stars  with initial  masses  on the  ZAMS  ranging from  1 to  $3.75\,
M_{\odot}$ (Althaus  et al. 2005;  Miller Bertolami \&  Althaus 2006).
All the  post-AGB evolutionary sequences were computed  using the {\tt
LPCODE}  evolutionary code  and were  followed through  the  very late
thermal pulse and the resulting  born-again episode that gives rise to
the  H-deficient, He-,  C-  and O-rich  composition characteristic  of
PG1159 stars.  The masses of  the resulting remnants are 0.530, 0.542,
0.556,  0.565,  0.589,  0.609,  0.664, and  $0.741\,  M_{\odot}$.   We
computed  $\ell= 1$  $g$-mode  adiabatic pulsation  periods for  these
equilibrium  models  with  the  same  numerical code  and  methods  we
employed in our previous works (C\'orsico \& Althaus 2006).

\begin{figure}[t]
\vskip 4cm
\includegraphics{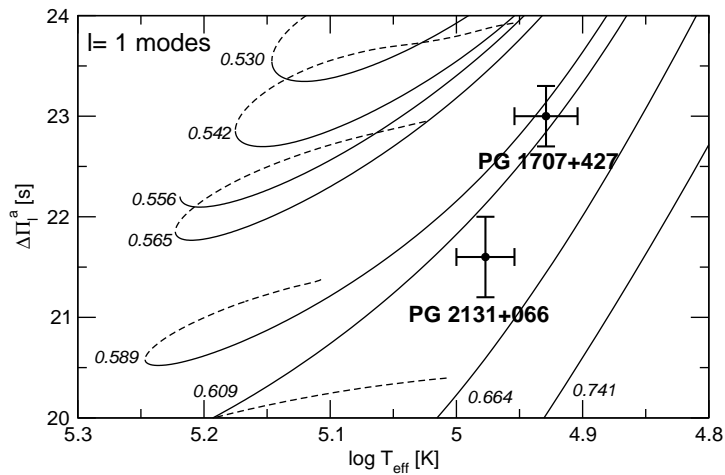}
\caption{\label{figure1}The  dipole   ($\ell=  1$)  asymptotic  period
         spacing in  terms of the effective  temperature for different
         stellar masses  (in solar units). Also shown  is the location
         of \pr\ and \pt.}
\end{figure}  

\section{Mass determination from the observed period spacing}  
  
\begin{figure}[t]  
\vskip 5cm
\includegraphics{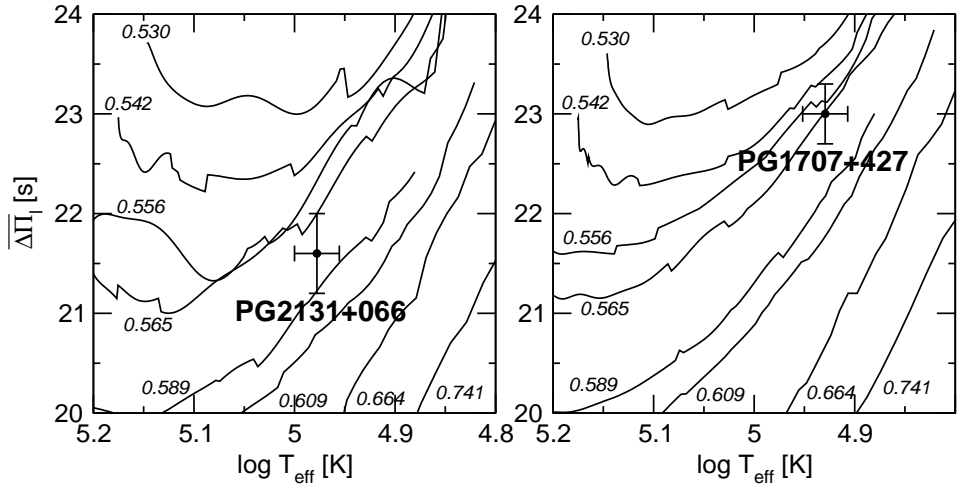}
\caption{\label{figure2}Same  as  Fig.   \ref{figure2},  but  for  the
         average of the computed period spacings.}  
\end{figure}

Here, we constrain the stellar mass  of \pr\ and \pt\ by comparing the
asymptotic  period spacing  and  the average  of  the computed  period
spacings with the {\sl observed} period spacing.  These approaches are
based on the fact that  the period spacing of PG1159 pulsators depends
primarily on  the stellar mass,  and the dependence on  the luminosity
and  the He-rich  envelope  mass fraction  is  negligible (Kawaler  \&
Bradley  1994;  C\'orsico   \&  Althaus  2006).   Fig.   \ref{figure1}
displays  the asymptotic  period  spacing  for $\ell=  1$  modes as  a
function of  the effective  temperature for different  stellar masses.
Also shown in this diagram is  the location of \pr, with $T_{\rm eff}=
95 \pm 5$ kK (Dreizler  \& Henber 1998) and $\Delta \Pi^{\rm O}_{\ell=
1}= 21.6 \pm 0.4$ s (Reed et al. 2000), and \pt, with $T_{\rm eff}= 85
\pm 5$ kK (Dreizler \&  Heber 1998) and $\Delta \Pi^{\rm O}_{\ell= 1}=
23.0 \pm 0.3$  s (Kawaler et al. 2004).  The asymptotic period spacing
is computed as in Tassoul  et al. (1990).  From the comparison between
the  observed $\Delta  \Pi^{\rm O}_{\ell=  1}$ and  $\Delta \Pi_{\ell=
1}^{\rm a}$  we found a stellar  mass of $0.627\,  M_{\odot}$ for \pr\
and $0.597\, M_{\odot}$  for \pt.  In Fig.  \ref{figure2}  we show the
run of average of the computed ($\ell= 2$) period spacings ($\ell= 1$)
for  \pr\  (left  panel), and  \pt\  (right  panel)  in terms  of  the
effective temperature  for all  of our PG1159  evolutionary sequences.
From the  comparison between  the observed $\Delta  \Pi^{\rm O}_{\ell=
1}$  and $\overline{\Delta \Pi_{\ell}}$,  we found  a stellar  mass of
$0.578  \,  M_{\odot}$  and  $0.566\,  M_{\odot}$ for  \pr\  and  \pt,
respectively.  These values  are $8.5 \%$ (for \pr)  and $5.5 \%$ (for
\pt) smaller than those derived through the asymptotic period spacing,
showing  once again  that  the asymptotic  approach overestimates  the
stellar mass  of PG1159 stars that,  like \pr\ and  \pt, exhibit short
and intermediate pulsation periods.
  
\section{Constraints from the individual observed periods}  
\label{fitting}  
  
In this approach  we seek pulsation models that  best matches the {\sl
individual} pulsation  periods of  \pr\ and \pt.   For both  stars, we
assume that all of the  observed periods correspond to $\ell= 1$ modes
(Kawaler et al. 1995; Kawaler et al. 2004).  The goodness of the match
between the  theoretical pulsation  periods ($\Pi_k^{\rm T}$)  and the
observed individual periods ($\Pi_i^{\rm  O}$) is measured by means of
a merit function, $\chi^2(M_*,T_{\rm  eff})$, which takes into account
the differences between the  observed and theoretical periods squared.
For  each star,  the PG  1159  model that  shows the  lowest value  of
$\chi^2$ will be adopted as  the ``best-fit model''.  For \pr, we find
a  clear seismological solution  corresponding to  a model  with $M_*=
0.589  \, M_{\odot}$  and  $T_{\rm eff}\approx  102$  kK.  This  model
provides an  excellent agreement between the  theoretical and observed
periods.   This is  quantitatively  reflected by  the  average of  the
absolute period  differences, $\overline{\delta \Pi_i}=  1.57$ and the
root-mean-square residual $\sigma_{_{\delta\Pi_i}}= 2.32$ s.  For \pt,
we find  a unambiguous seismological  solution for a model  with $M_*=
0.542  \, M_{\odot}$ and  $T_{\rm eff}\approx  89.5$ kK,  an effective
temperature compatible with the spectroscopic determination.  We found
$\overline{\delta \Pi_i}= 1.75$ s and $\sigma_{_{\delta \Pi_i}}= 1.99$
s.
  
The main  features of  our best-fit model  for \pr\ are  summarized in
Table \ref{table3}.  The total mass of the best-fit model ($M_*= 0.589
\, M_{\odot}$) is in agreement with the value derived from the average
of the computed  period spacings ($M_* \sim 0.578  \, M_{\odot}$), but
at odds ($\sim  6 \%$ smaller) with that  inferred from the asymptotic
period spacing  ($M_*= 0.627\, M_{\odot}$).  Also, the  $M_*$ value of
our best-fit model is substantially larger than the spectroscopic mass
of  $0.53\pm  0.1  \,  M_{\odot}$  (Dreizler  \&  Heber  1998;  Miller
Bertolami  \&  Althaus  2006)  for  \pr.  A  discrepancy  between  the
seismological  and  the spectroscopic  values  of  $M_*$ is  generally
encountered among PG1159 pulsators  (C\'orsico et al. 2007a; C\'orsico
et al.  2007b).  Until  now, the seismological  mass of \pr\  has been
about $11 \%$ larger ($\Delta  M_* \approx 0.06\, M_{\odot}$) than the
spectroscopic  mass  if  we  consider  the early  estimation  for  the
seismological  mass quoted in  Reed et  al. (2000).   In light  of the
best-fit  model derived in  this paper,  this discrepancy  is slightly
reduced to about $7 \%$ ($\Delta M_* \approx 0.04\, M_{\odot}$).
  
\begin{table}  
\begin{center}  
\caption{The main  characteristics of \pr.  Column  (1) corresponds to
         spectroscopic results  from Dreizler \&  Heber (1998), column
         (2) presents   results  from  previous   pulsational  studies
         (Kawaler  et al.  1995;  Reed  et al.  2000)  and column  (3)
         provides  the results  from the  seismological model  of this
         work, respectively.}
\begin{tabular}{lccc}  
\br
                       & (1)                   & (2)               &  (3)                      \\  
\mr
$T_{\rm eff}$ [kK]     & $95 \pm 5$            & ---               & $102.18_{-2.8}^{+3.0}$    \\  
$M_*$ [$M_{\odot}$]    & $0.55\pm 0.1$         & $0.608 \pm 0.01$  & $0.589_{-0.024}^{+0.020}$ \\   
$\log g$ [cm/s$^2$]    & $7.5 \pm 0.5$         &        ---        & $7.63_{-0.14}^{+0.12} $   \\   
$\log (L_*/L_{\odot})$ & $1.6 \pm 0.3$         & $1.41 \pm 0.5$    & $1.57_{-0.06}^{+0.07}$    \\    
$\log(R_*/R_{\odot})$  & ---                   & $-1.73$           & $-1.71_{-0.05}^{+0.06}$   \\    
$d$  [pc]              & $681_{-137}^{+170}$   & $668_{-83}^{+78}$ & $830_{-224}^{+300}$       \\   
$\pi$ [mas]            & $1.47_{-0.3}^{+0.4}$  & $1.50\pm 0.2$     & $1.2_{-0.3}^{+0.4}$       \\   
\br
\end{tabular}  
\label{table3}  
\end{center}  
\end{table}  
  
The main properties of our best-fit  model for \pt\ are shown in Table
\ref{table4}.  The stellar mass of our best-fit model is $M_*= 0.542\,
M_{\odot}$, in  agreement with the  value derived from the  average of
the computed  period spacings ($M_*  \sim 0.566\, M_{\odot}$),  but at
odds ($\sim 9 \%$ lower) with that inferred from the asymptotic period
spacing ($M_*= 0.597  \, M_{\odot}$). On the other  hand, we note that
$M_*$  for the  best-fit  model  is in  excellent  agreement with  the
spectroscopic  derivation  ($0.542   \,  M_{\odot}$  versus  $0.53  \,
M_{\odot}$).  Until now, the seismological  mass of \pt\ has been more
than $7  \%$ larger ($\Delta  M_* \approx 0.04\, M_{\odot}$)  than the
spectroscopic mass  if we adopt  for the seismological mass  the value
found  previously (Kawaler  et al.  2004).  In  light of  our best-fit
model, this discrepancy  is strongly reduced to about  $2 \%$ ($\Delta
M_* \approx 0.012 \, M_{\odot}$).
  
\begin{table}  
\begin{center}  
\caption{Same as Table \ref{table3},  but for \pt. Column (2) presents
         results from the pulsational study of Kawaler et al. (2004).}
\begin{tabular}{lccc}
\br
                       & (1)                   & (2)             & (3)                       \\
\mr  
$T_{\rm eff}$ [kK]     & $85 \pm 4.5$          & ---             & $89.5_{-1.8}^{+1.7}$      \\  
$M_*$ [$M_{\odot}$]    & $0.53\pm 0.1$         & $0.57 \pm 0.02$ & $0.542_{-0.012}^{+0.014}$ \\   
$\log g$ [cm/s$^2$]    & $7.5 \pm 0.3$         & ---             & $7.53_{-0.08}^{+0.09}$    \\   
$\log (L_*/L_{\odot})$ & ${1.4 \pm 0.3}$       & $1.36$          & $1.40\pm 0.04$            \\    
$\log(R_*/R_{\odot})$  & ---                   & ---             & $-1.68\pm 0.04$           \\    
$d$  [pc]              & $1300_{-300}^{+1000}$ & ---             & $730_{-175}^{+230}$       \\   
$\pi$ [mas]            & $0.77_{-0.3}^{+0.2}$  & ---             & $1.4\pm0.4$               \\   
\br
\end{tabular}  
\label{table4}  
\end{center}  
\end{table}  
  
\section{Summary and conclusions}  
\label{conclusions}  
  
In this  work we have estimated the  stellar mass of \pr\  and \pt\ on
the basis  of the period-spacing  information alone, and we  have also
been successful  in finding seismological  models for both  stars from
period-to-period fits.   The present study closes our  short series of
seismological  studies  of  PG1159  stars  based on  a  grid  of  full
evolutionary  models  characterized  by  thick He-rich  envelopes,  as
dictated by canonical stellar evolution.
    
\ack
  
Part of this work was supported by PIP 6521 grant from CONICET, by MEC  
grant AYA05--08013--C03--01, by the  European Union FEDER funds and by  
the AGAUR.  

\section*{References}

\begin{thereferences}
  
\item Althaus L G Serenelli A M Panei et al. 2005 {\sl A\&A} {\bf 435}
      631
\item C\'orsico A H \& Althaus L G 2006 {\sl A\&A} {\bf 454} 863
\item C\'orsico A H Althaus L G Miller Bertolami M M \& Werner K 2007a
      {\sl A\&A} {\bf 461} 1095
\item C\'orsico  A H Miller  Bertolami M M  Althaus L G Vauclair  G \&
      Werner K 2007b {\sl A\&A} {\bf 461} 1095
\item C\'orsico  A H  Althaus L G  Kepler S  O Costa J  E S  \& Miller
      Bertolami M M 2008 {\sl A\&A} {\bf 478} 869
\item Costa J E  S Kepler S O Winget D E et  al.  2008 {\sl A\&A} {\bf
      477} 627
\item Dreizler S \& Heber U 1998 {\sl A\&A} {\bf 334} 618
\item Fu J N Vauclair G Solheim,  J E et al. 2007 {\sl A\&A} {\bf 467}
      237
\item Kawaler S D \& Bradley P A 1994 {\sl ApJ} {\bf 427} 415
\item Kawaler S D  O'Brien M S Clemens J C et  al. 1995 {\sl ApJ} {\bf
      450} 350
\item Kawaler S D Potter E M  Vuckovi\'c M et al. 2004 {\sl A\&A} {\bf
      428} 969
\item Miller  Bertolami M M \& Althaus  L G 2006 {\sl  A\&A} {\bf 454}
      845
\item Reed M D Kawaler S D \& O'Brien M S 2000 {\sl ApJ} {\bf 545} 429
\item Tassoul M Fontaine G \& Winget D E 1990 {\sl ApJS} {\bf 72} 335
\item Vauclair  G Moskalik P  Pfeiffer B et  al. 2002 {\sl  A\&A} {\bf
      381} 122
\item Werner K \& Herwig F 2006 {\sl PASP} {\bf 118} 183
\item Winget D E \& Kepler S O 2008, {\sl ARA\&A} {\bf 46} 157
 
\end{thereferences}

\end{document}